\documentclass[aps,prl,floatfix,twocolumn,showpacs,nofitinbib,superscriptaddress,hyperref=pdftex,citeautoscript]{revtex4-2}

\usepackage{soul}
\usepackage{titletoc}

\titlecontents{section}%
  [3em]
  {\addvspace{1em plus 0pt}\bfseries}
  {\contentslabel{3em}}
  {\hspace{-3em}}
  {\hfill\contentspage}
  [\addvspace{0pt}]

\titlecontents{subsection}%
  [5em]
  {\addvspace{.5em plus 0pt}}
  {\contentslabel{3em}}
  {\hspace{-3em}}
  {\hfill\contentspage}
  [\addvspace{0pt}]

\titlecontents{subsubsection}%
  [7em]
  {\addvspace{.5em plus 0pt}}
   {\contentslabel{4em}}
  {\hspace{-4em}}
  {\hfill\contentspage}
  [\addvspace{0pt}]
  
\usepackage{lmodern}
\usepackage{epsfig}
\usepackage{amsmath}
\usepackage{amssymb}
\usepackage{amsfonts}
\usepackage{dsfont}
\usepackage{comment,color}
\usepackage{lipsum}
\usepackage{hyperref,mathtools}
\usepackage{makecell}
\usepackage{todonotes}
\usepackage{array}
\usepackage{physics}
\usepackage{xcolor,float}
\usepackage[sep=3pt, offset=1em]{simpler-wick}
\usepackage{tikz}
\usetikzlibrary{quantikz}
\usepackage{lipsum}
\usepackage{siunitx}
 \usepackage{orcidlink}
\usepackage{tikz}
\usepackage{blkarray}

\hypersetup{colorlinks=true,linkcolor=blue,anchorcolor=blue,citecolor=blue,filecolor=blue,urlcolor=blue,bookmarksnumbered=true,pdfview=FitB}

\newcommand{\beqa}{\begin{eqnarray}}
\newcommand{\eeqa}{\end{eqnarray}}

\newcommand{\ua}{\uparrow}
\newcommand{\da}{\downarrow}

\newcommand{\T}{\mathcal T}
\renewcommand{\bm}{\boldsymbol}
\def\S{\mathcal S_{\mathrm{PH}}}

\usepackage[normalem]{ulem}
\renewcommand\sout{\bgroup\markoverwith{\textcolor{red}{\rule[0.5ex]{2pt}{0.4pt}}}\ULon}

\begin{document}

\hsize\textwidth\columnwidth\hsize\csname@twocolumnfalse\endcsname
\def\titlecommand{Fermionic Quantum Simulation on Andreev Bound State Superlattices}
\title{\titlecommand}

\author{Peter D. Johannsen\orcidlink{0000-0003-4570-8881}}
\affiliation{Center for Quantum Devices, Niels Bohr Institute, University of Copenhagen, 2100 Copenhagen, Denmark}
\author{Constantin Schrade}
\affiliation{Hearne Institute of Theoretical Physics, Department of Physics \& Astronomy, Louisiana State University, Baton Rouge LA 70803, USA}

\date{\today}

\vskip1.5truecm
\begin{abstract}
Arrays of superconducting qubits and cavities offer a promising route for realizing highly controllable artificial materials. However, many analog simulations of superconducting circuit hardware have focused on bosonic systems. Fermionic simulations, on the other hand, have largely relied on digital approaches that require non-local qubit couplings, which could limit their scalability. Here, we propose and study an alternative approach for analog fermionic quantum simulation based on arrays of \textit{coherently coupled} mesoscopic Josephson junctions. These Josephson junction arrays implement an effective superlattice of Andreev bound state ``atoms" that can trap individual fermionic quasiparticles and, due to their wavefunction overlap, mediate quasiparticle hoppings. By developing a Wannier function approach, we show that these Andreev bound state arrays form an all-superconducting and circuit QED-compatible platform for emulating lattice models of fermionic quasiparticles that are phase- and gate-programmable. Interestingly, we also find that the junction lattices can undergo a topological transition and host fermionic boundary modes that can be probed by conductance measurements. We hope our results will inspire the realization of artificial and possibly topological materials on Andreev bound state quantum simulators.\end{abstract}

\maketitle
\textit{Introduction.} Superconducting qubits are a leading modality for intermediate-scale~\cite{preskill2018quantum}, error-corrected~\cite{andersen2020repeated,marques2022logical,krinner2022realizing,google2023suppressing} quantum information devices. Among their most attractive features is their flexible electromagnetic tunability, which not only enabled high-fidelity
gates~\cite{sung2021realization,ding2023high,zhang2023tunable} but also opened the way to highly-controllable artificial matter setups on superconducting qubit and cavity arrays~\cite{koch2010time,underwood2012low,houck2012chip,anderson2016engineering,carusotto2020photonic,yanay2020two,karamlou2022quantum,braumuller2022probing}. 
However, despite notable progress in the \textit{analog} superconducting quantum simulations field, most research has focused on emulating 
bosonic systems due to the inherent bosonic nature of superconducting circuits. In contrast, simulating fermionic systems has been primarily restricted to \textit{digital} simulation approaches~\cite{barends2015digital,arute2020observation}, limited by moderate system sizes because of long-range couplings that arise from fermion-to-qubit mappings~\cite{havlivcek2017operator}. Hence, there is a current need for discovering efficient ways to simulate fermionic systems with superconducting circuits.

Recently, a new frontier for artificial superconducting arrays has opened up, enabled by regular patterning of semiconductors with superconductors~\cite{bottcher2018superconducting,bottcher2022dynamical}. 
Such hybrid lattices have already shown intriguing features, such as superconductor-insulator~\cite{bottcher2018superconducting} and dynamical vortex transitions~\cite{bottcher2022dynamical}. An additional property
of these systems is the formation of Andreev bound states (ABSs)~\cite{beenakker1991universal,martinis2004superconducting,sauls2018andreev,kornich2022andreev,li2023anomalous,hinderling2023flip,hinderling2024direct}, which arise in the Josephson junction (JJ) region when superconductors encapsulate a semiconductor. Such ABSs have been widely studied for novel superconducting qubit designs~\cite{zazunov2003andreev,chtchelkatchev2003andreev,padurariu2010theoretical,park2017andreev,hays2020continuous,cerrillo2021spin,pita2023direct,pita2023strong,ackermann2023dynamical,hays2021coherent,matute2022signatures,schrade2018majorana,schrade2022quantum,larsen2020parity,schrade2022protected,maiani2022entangling,ciaccia2024charge} and Josephson diode applications~\cite{zazunov2009anomalous,brunetti2013anomalous,baumgartner_supercurrent_2022,lotfizadeh2023superconducting,gupta2023gate,costa2023microscopic,zhang2022evidence,davydova_universal_2022,banerjee2023phase,zazunov2023nonreciprocal,zazunov2023approaching,pal2022josephson,souto2022josephson,ciaccia2023gate,valentini2023parity,souto2024tuning,legg2023parity,bozkurt2023double,trahms2023diode,maiani2023nonsinusoidal,hess2023josephson,matsuo2023engineering,matsuo2023josephson,coraiola2023flux}. Importantly, ABSs also have the unique capability of trapping fermionic quasiparticles. An interesting question is if this \textit{fermionic} quasiparticle degree of freedom can be exploited for analog quantum simulation in an \textit{all-superconducting} architecture?

\begin{figure}[!t]
    \centering
    \includegraphics[width=\linewidth]{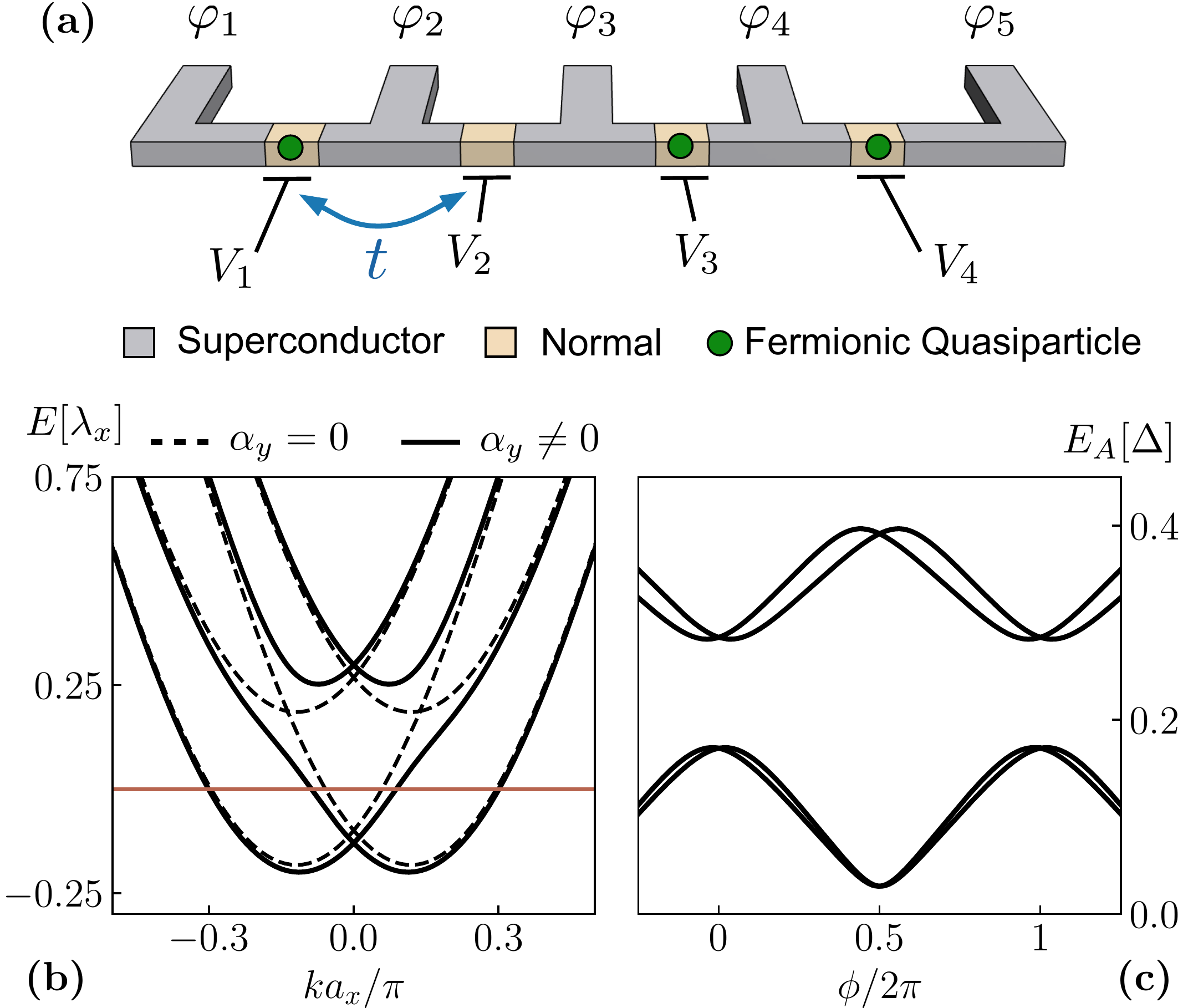}
    \caption{
    (a) Array of \textit{coherently coupled} JJs with phase- and voltage-biases, \{$\varphi_{i},V_{i}$\}, realizing a quantum simulator for lattice models of fermionic quasiparticles. 
    (b) Normal-state bandstructure in the presence (solid) and absence (dashed) of a transversal spin-orbit coupling, $\alpha_{y}$.  
    (c) ``Atomic'' ABS energy levels of a single JJ versus the phase drop, $\phi$, showing spin-orbit-induced spin-splittings \cite{Supplemental}.}
    \label{fig:1}
\end{figure}

In this work, we address this question by introducing lattices of \textit{coherently coupled} mesoscopic JJs as a route to analog fermionic quantum simulations with superconducting circuits. 
In such JJ lattices, ABSs localized in the JJ regions take on the role of ``enlarged atomic orbitals'' that span nanometer-scale distances over many microscopic sites. 
Furthermore, the overlapping of neighboring ABS wavefunctions leads to effective ``inter-atomic bondings'', facilitating the hopping of fermionic quasiparticles on the JJ lattice. Leveraging the gate- and phase-programmability of the bonds, we show that such JJ lattices form a versatile, all-superconducting platform for simulating lattice models of fermionic quasiparticles. To illustrate our approach, we consider a minimal 
$1d$ JJ array, as shown in Fig.\,\ref{fig:1}(a), and use a Wannier function approach to develop an effective tight-binding model of fermionic quasiparticles with a gate- and phase-programmable band structure. 
Notably, by in-situ tuning of the phases and gates, we find that this system can undergo a topological phase transition and host fermionic boundary modes that can be probed by conductance measurements. 
We hope that our results will inspire explorations of ABS superlattices as a novel platform for simulating artificial and topological materials.

\textit{Setup.}
Our proposed quantum simulator setup is shown in Fig.\,\ref{fig:1}(a). It comprises an array of $N$ mesoscopic JJs, each hosting ABSs localized in the JJ regions. For modeling this setup, we consider the following tight-binding model~\cite{hays2021coherent,matute2022signatures}, 
\begin{equation}
\begin{split}
H_{}
&=
\sum_{i,j,s}(\varepsilon_{ij}-\mu)\,c^{\dagger}_{i,j,s}c_{i,j,s}
+
\lambda_{x}\,c^{\dagger}_{i+1,j,s}c_{i,j,s}
\\
&
+
s
\alpha_{x}\,c^{\dagger}_{i+1,j,s}c_{i,j,\bar{s}}
+
\sum_{i,j}\Delta_{i}e^{i\varphi_{i}}\,c_{i,j,\downarrow}c_{i,j,\uparrow}
\\
&+\sum_{i,j,s}\lambda_{y}\,c^{\dagger}_{i,j,s}c_{i,j+1,s}
+
i\alpha_{y}\,c^{\dagger}_{i,j,s}c_{i,j+1,\bar{s}}
+
\text{H.c.}
\end{split}
\label{Eq1}
\end{equation}
Here, $c_{i,j,s}$ is the spin-$s$ electron annihilation operator for a site $(i,j)$. The on-site energies, $\varepsilon_{ij}$, are measured relative to the chemical potential, $\mu$. The hoppings in the longitudinal and transversal directions are given by $\lambda_{x,y}$, and the corresponding spin-orbit couplings are $\alpha_{x,y}$. The magnitude and phase of the superconducting order parameter are $\Delta_{i}$ and $\varphi_{i}$. For simplicity, we set $\Delta_{i}\equiv\Delta$ and $\varphi_{i}\equiv\varphi_{n}$ if the site $(i,j)$ is on the $n^{\text{th}}$ superconducting grain and $\Delta_{i}\equiv0$ if $(i,j)$ is a site in a JJ region. We highlight that Eq.\,\eqref{Eq1} describes a JJ array for a superconductor-semiconductor structure. However, our results will apply also to other ABS systems~\cite{schindler2018higher}. 

\textit{Andreev ``atom" and ``molecule".}
To start, it is instructive to recapitulate the fundamental components of our setups, the ABS ``atom'' ($N=1$) and ``molecule''~\cite{su2017andreev,pillet2019nonlocal,kornich2019fine,junger2023intermediate,matsuo2023phase,haxell2023demonstration,pillet2023josephson,van2023charge,bordin2024supercurrent,coraiola2023phase} ($N=2$). For simplicity, we will thereby focus on the case of two transversal sites ($j=1,2$)
, which qualitatively reproduces the ABS spectrum of nanowire JJs~\cite{hays2021coherent,matute2022signatures}. 

For the case of $N=1$ JJ, we initially consider the normal state spectrum ($\Delta\equiv 0$), see Fig.\,\ref{fig:1}(b). It consists of parabolic bands that shift horizontally due to the longitudinal spin-orbit coupling, $\alpha_{x}$, and undergo avoided crossings due to the transversal spin-orbit coupling, $\alpha_{y}$. Because of these avoided crossings, the Fermi velocities for the left- and right-moving modes in the JJs are generally unequal. In the superconducting state, these unequal Fermi velocities lead to a spin-splitting of the ABS levels, as shown in Fig.\,\ref{fig:1}(c).

\begin{figure}[!b]
    \centering
    \includegraphics[width=\linewidth]{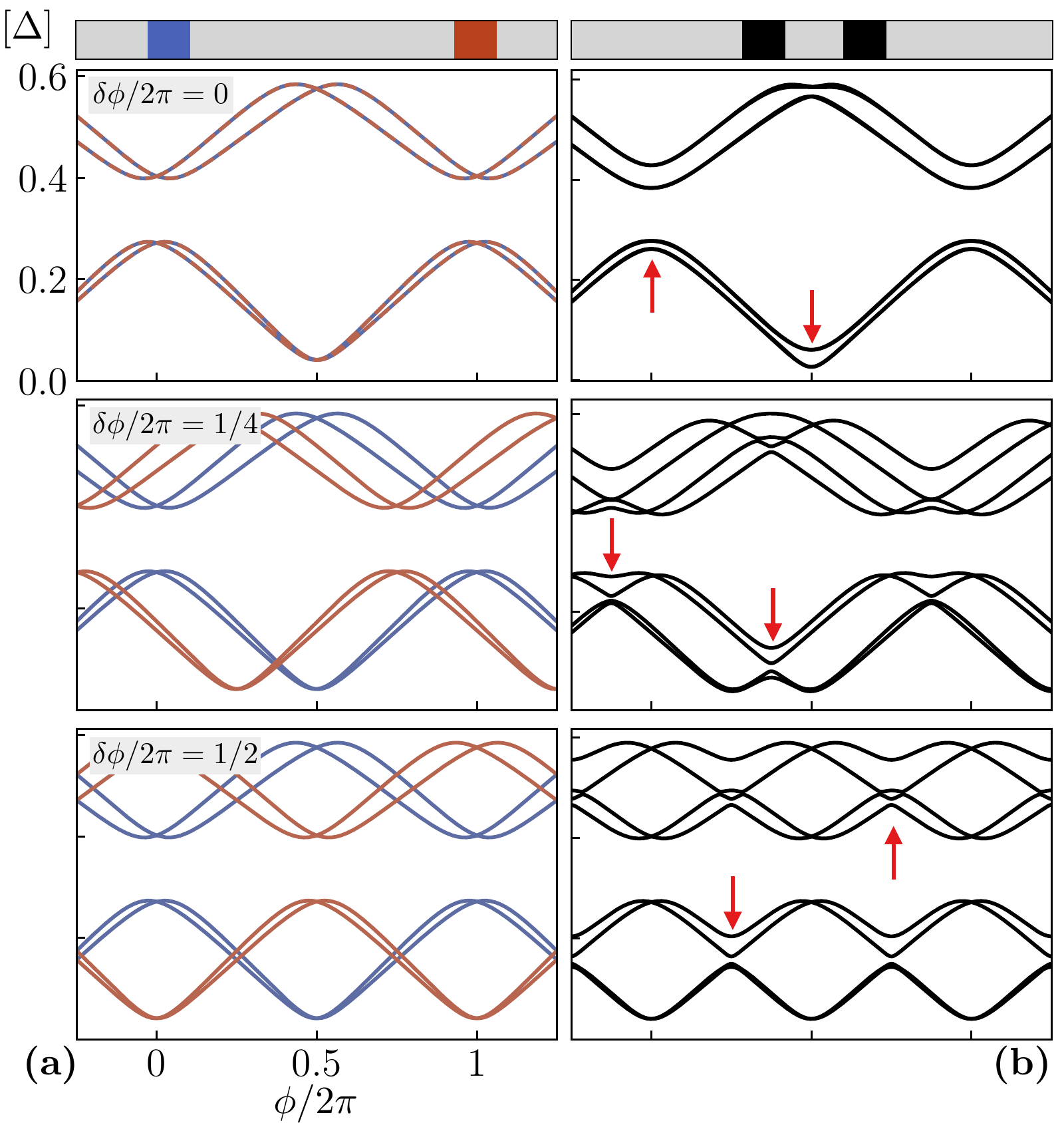}
    \caption{(a) ABS levels versus $\phi=\phi_{1}$ for various $\delta\phi\equiv\phi_{2}-\phi_{1}$ and large JJ separation ($d\gg\xi_{S}$). The ABS wavefunction overlap is negligible. Hence, there is no coupling between the left (blue) and right (red) JJ. (b) Same as (a) but for $d\lesssim\xi_{S}$. The ABS wavefunctions overlap, inducing hybridization gaps at which fermionic quasiparticle hoppings are enhanced.}

    \label{fig:2}
\end{figure}
Next, we proceed to $N=2$ JJs and first examine the situation when the JJ separation, $d$, is much larger than the superconducting coherence length, $d\gg\xi_{S}$. In this case, we expect that the wavefunction overlap of ABSs in the two JJs is negligibly small, and the JJs can be treated as independent. To confirm this, we calculate the ABS levels as a function of $\phi = \phi_1$ for various $\delta\phi=\phi_2 - \phi_1$, as shown in Fig.\,\ref{fig:2}(a), assuming the same on-site energies in the two JJs for simplicity. When $\delta\phi=0$, the ABS levels are identical to the $N=1$ case due to identical phase drops across the JJs. In contrast, when $\delta\phi\neq0$, the ABS levels for each JJ shift relative to each other due to the finite offset between the phase drops. Crucially, though, no hybridizations appear between the spectrum branches of the two JJs.  This aligns with our expectation of negligible ABS wavefunction overlap.

The situation of two distant JJs should now be compared to JJs with a separation that approaches the superconducting 
coherence length, $d\lesssim\xi_{S}$. As shown in Fig.\,\ref{fig:2}(b), the ABS levels look identical to the previous case but with the notable exception that band gaps open at the crossing points of the branches that correspond to the left and right JJ. Near these crossings, the ABS wavefunctions hybridize. In particular, localized states in each JJ are no longer system eigenstates. Instead, a coupling emerges between states in both JJs, facilitating the hopping of fermionic quasiparticles.

\textit{Wannier wavefunctions.} The hopping of fermionic quasiparticles is the all-important ingredient of our quantum simulator platform. To explicitly compute the hopping amplitudes, we now introduce an approach to construct Wannier wavefunctions that are localized in the two JJs but are not eigenstates when $d\lesssim\xi_{S}$. The construction of such Wannier wavefunctions for our ABS superlattice will make use of a ``projection method'', similar to the construction of Wannier wavefunctions for ordinary crystals \cite{vanderbilt_2018,RevModPhys.84.1419} or moir\'{e} superlattices \cite{zhang2019bridging}. 

\begin{figure}[!t]
    \centering
    \includegraphics[width=\linewidth]{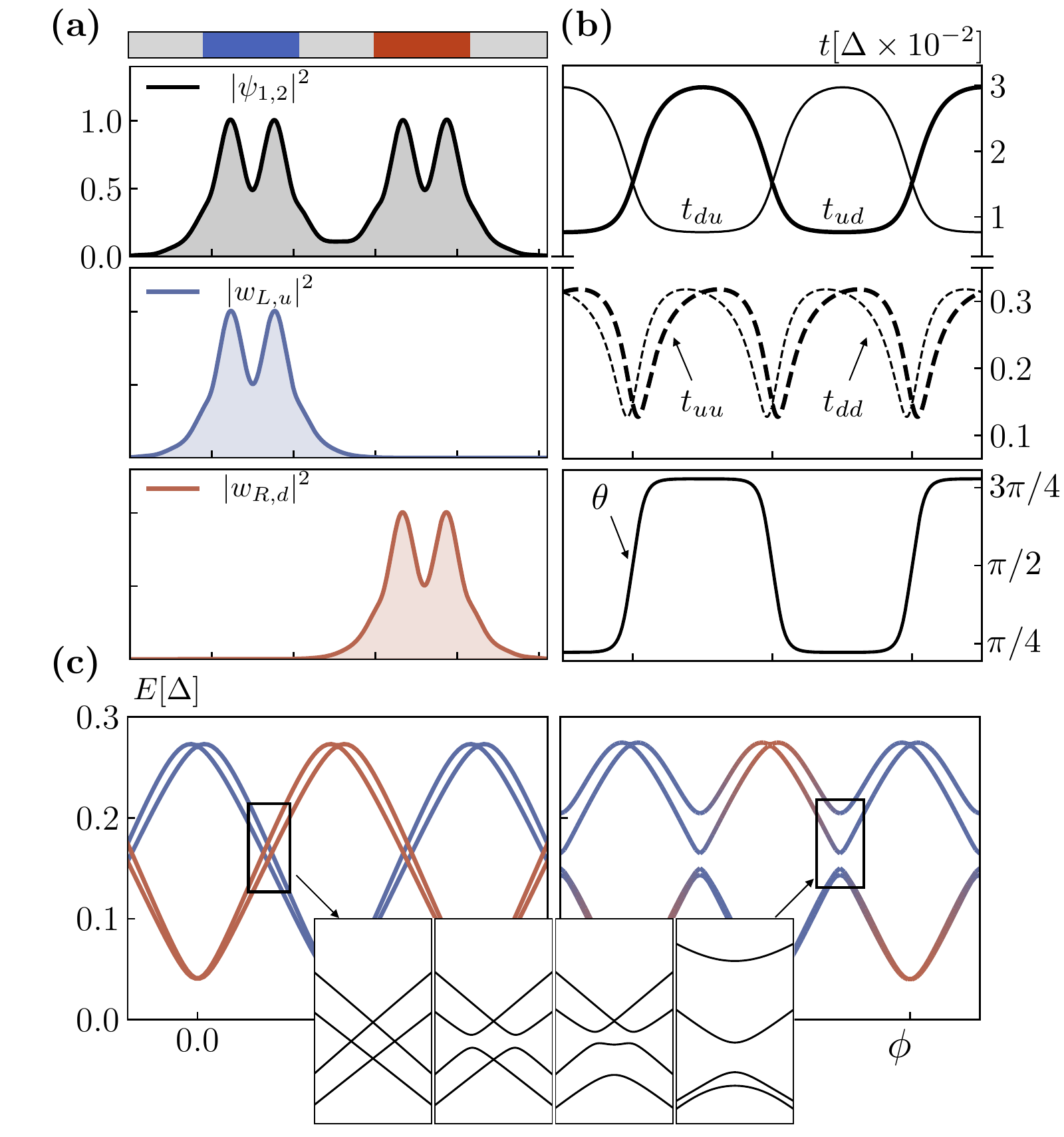}
    \caption{
    (a) 
    Lowest-energy eigenfunctions (black), $|\psi_{1,2}|^{2}$, and related Wannier wavefunctions (blue,red), $|w_{n,\sigma}|^{2}$, at $\delta\phi \equiv \phi_{2}-\phi_{1}= \pi$ and $\phi\equiv\phi_{1}= \pi/2$. (b) Hopping amplitudes, $t_{\sigma\sigma'}$, and phase, $\theta = (\theta_{uu} + \theta_{dd} - \theta_{ud}-\theta_{du})/2$ that is gauge-invariant for $N=2$ JJs, versus $\phi$ for $\delta\phi=\pi$. (c) Effect of the hoppings on the ABS levels. Zoom-ins show
    gap openings when sequentially turning on the hoppings, \{$t_{uu,dd},t_{du},t_{ud}$\}
    }
    \label{fig:3}
\end{figure}

To explain our method, we focus on the four lowest energy states of the $N=2$ system, $|\psi_{1,\dots,4}\rangle$, and introduce 
a corresponding set of four localized ``trial states'', $|g_{n,\sigma}\rangle$. These trial states are obtained from the two lowest energy eigenstates ($\sigma=u,d$) of each JJ ($n=1,2$) in the uncoupled limit when $d\gg\xi_{S}$. The trial states are then projected onto the eigenstates of the coupled JJ system. This procedure yields new states,  $|g'_{n,\sigma}\rangle=\kappa_{1,n\sigma}|\psi_{1}\rangle+\kappa_{2,n\sigma}|\psi_{2}\rangle+\kappa_{3,n\sigma}|\psi_{3}\rangle+\kappa_{4,n\sigma}|\psi_{4}\rangle$ with $\kappa_{\ell,n\sigma}=\langle \psi_{\ell}|g_{n,\sigma}\rangle$, which are localized in the JJ regions but are not yet orthonormal. To orthonormalize the states, we use a singular-value decomposition, which results in four orthonormal Wannier states, $|w_{n,\sigma}\rangle$. 
The resulting Wannier wavefunctions, $w_{n,\sigma}(i,j)=\langle i,j|w_{n,\sigma}\rangle$, are depicted in Fig.\,\ref{fig:3}(a) and allow us to introduce new fermionic operators, describing quasiparticles with effective spin $\sigma$ in the $n^{\text{th}}$ JJ, 
\begin{equation}
f_{n,\sigma}=\sum_{i,j}\boldsymbol{w}_{n,\sigma}(i,j)^{T}\cdot\boldsymbol{c}_{i,j}.\label{main:eq:fdef}
\end{equation}
Here, $\boldsymbol{c}_{}=(c_{\ua},c_{\da},c^{\dag}_{\da},-c^{\dag}_{\ua})^{T}$ and $\boldsymbol{w}_{}=(w_{\ua},w_{\da},\tilde{w}_{\da},\tilde{w}_{\ua})^{T}$. We emphasize that in the above expression, $(i,j,s)$ denote degrees of freedom on the \textit{microscopic} lattice, while $(n,\sigma)$ denote the emergent degrees of freedom of the \textit{macroscopic} ABS superlattice. Lastly, as the localized fermionic operators in Eq.\,\eqref{main:eq:fdef} are related by a unitary to the Bogoliubov eigenoperators of the coupled JJ system, we note that they obey the standard anti-commutations, $\{f_{n,\sigma},f^{\dag}_{n',\sigma'}\}=\delta_{nn'}\delta_{\sigma\sigma'}$ and $\{f_{n,\sigma},f_{n',\sigma'}\}=0$\,\cite{Supplemental}. 

\textit{Effective Hamiltonian.}
We now proceed by projecting the Wannier states onto the tight-binding Hamiltonian of Eq.\,\eqref{Eq1}. This procedure yields our quantum simulator Hamiltonian, i.e., the effective Hamiltonian of fermionic quasiparticles hopping on the ABS superlattice,
\begin{equation}
\begin{split}
H^{}_{\text{eff}} 
&= 
\sum_{n,\sigma}
\epsilon_{n,\sigma}\, f^{\dag}_{n,\sigma}f_{n,\sigma}
+
\sum_{\langle n,n' \rangle}
\sum_{\sigma\sigma'} 
t^{nn'}_{\sigma\sigma'}\,
f^{\dag}_{n,\sigma}f_{n',\sigma'}
\end{split}\label{Eq:Heff}
\end{equation}
Here, the on-site energies and nearest-neighbor hoppings, $\epsilon_{n,\sigma}$ and $t^{nn'}_{\sigma\sigma'}=(t^{n'n}_{\sigma\sigma'})^{*}$, are given by, 
\begin{equation}
\begin{split}
\epsilon_{n,\sigma}&=
\sum_{i,j}\boldsymbol{w}_{n,\sigma}(i,j)^{\dag}\cdot
\mathcal{H}_{}\cdot\boldsymbol{w}_{n,\sigma}(i,j),
\\
t^{nn'}_{\sigma\sigma'}&=
\sum_{i,j}\boldsymbol{w}_{n,\sigma}(i,j)^{\dag}\cdot
\mathcal{H}_{}\cdot\boldsymbol{w}_{n',\sigma'}(i,j),
\label{hoppings}
\end{split}
\end{equation}
with $\mathcal{H}_{}$ denoting the Nambu space matrix form of $H_{}$. 

The effective Hamiltonian in Eq.\,\eqref{Eq:Heff} and the expressions for the on-site energies and hoppings in Eq.\,\eqref{hoppings} are a central result of our work. 
They provide an explicit mapping between the \textit{microscopic} lattice and the \textit{macroscopic} ABS superlattice, thereby enabling quantum simulation of fermionic lattice models. Moreover, Eq.\,\eqref{hoppings} also provides an approach for computing the on-site energies and hoppings from the Wannier functions, see Fig.\,\ref{fig:3}(b)\,\cite{kwant}. In particular, we find that the on-site energies, $\epsilon_{n,\sigma}$, match the ABS levels in the previously discussed limit of uncoupled JJs, $d\gg\xi_{S}$. Similarly, as shown in Fig.\,\ref{fig:3}(c), the hoppings, $t^{nn'}_{\sigma\sigma'}$, induce the energy gaps through ABS hybridization when $d\lesssim \xi_{S}$, in agreement with our earlier considerations.

\begin{figure*}[!t]
     \includegraphics[width=0.9\textwidth]{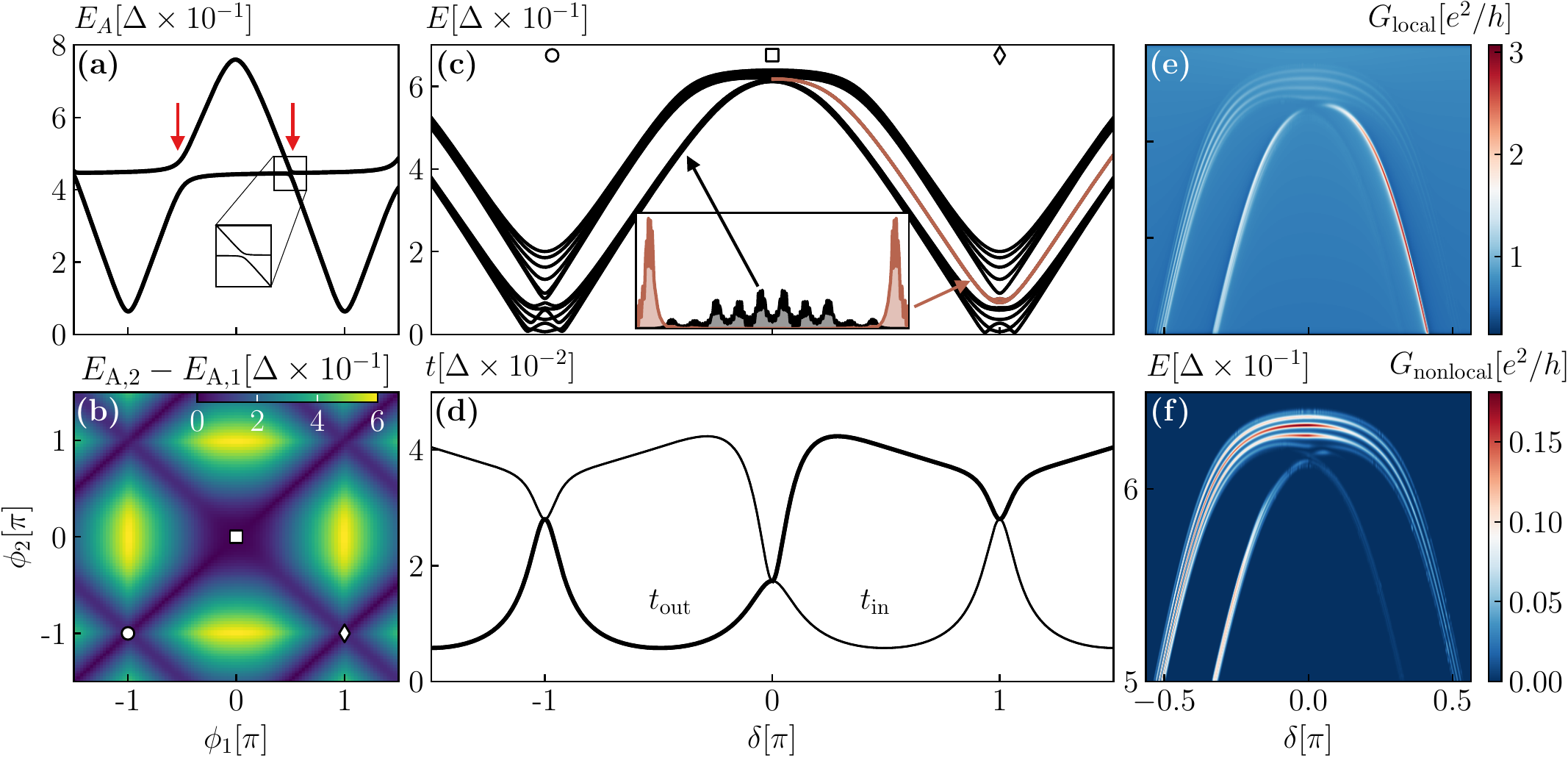}
     \caption{(a) ABS levels for $N=2$ JJs versus $\phi_{1}$ and for $\phi_{2}=\pi/2$. The two anti-crossings (arrows) with different magnitudes can be used to create the staggered hybridization required for simulating a fermionic SSH model. (b) Separation of the ABS levels from (a) versus $(\phi_{1},\phi_{2})$. When moving along the path $(\phi_{1},\phi_{2})=(-\pi,-\pi)$  $\rightarrow$ $(0,0)$ $\rightarrow$ $(\pi,-\pi)$ (indicated by the circle, square, and diamond), the relative magnitude of the two anticrossing reverses, a feature that can be used phase-tune between a trivial and topological SSH model. (c) ABS levels for a $N=12$ JJ lattice, when moving from a trivial phase configuration, $\bm\phi_{\text{triv}}\equiv\{\phi_{*},\phi_{*},-\phi_{*},-\phi_{*},\cdots\}$,
     to a topological phase configuration, $\bm\phi_{\text{topo}}\equiv\{\phi_{*},-\phi_{*},-\phi_{*},\phi_{*},\cdots\}$, via the interpolation $\bm\phi(\delta)$ given in the text. Topological bound states (red) appear for $\delta>0$. (d) Corresponding values of the staggered hoppings, $t_{\text{out}}$ and $t_{\text{in}}$. (e-f) Local and non-local conductance when connecting the JJ array on both ends to normal metal leads. The local conductance shows the emergence of the topological bound state levels.}\label{fig:4}
\end{figure*}

\textit{Topological lattice models}. 
As a next step, we will demonstrate the utility of our Wannier function approach by applying it to a paradigmatic $1d$ topological lattice system: the \textit{fermionic} Su-Schrieffer-Heeger (SSH) model. The Hamiltonian for the SSH model is given by,
\begin{equation}
H^{(\text{SSH})}=\sum_{n\, \text{even}}  t_\mathrm{in}f^\dagger_{n}f_{n+1}+\sum_{n\, \text{odd}}t_\mathrm{out}f^{\dagger}_{n}f_{n+1} + \mathrm{H.c}, \label{eq:SSH}
\end{equation}
where ($t_\mathrm{in},t_\mathrm{out}$) are staggered hopping amplitudes. Despite its simplicity, the SSH model provides a fundamental example of the realization of topological edge modes when $t_\mathrm{out}>t_\mathrm{in}$ and has been experimentally realized across various platforms~\cite{atala2013direct,meier2016observation,st2017lasing,de2019observation,cai2019observation,kiczynski2022engineering}. Here, we will simulate the fermionic SSH model in an array of \textit{equally spaced} JJs, where the superconducting phases permit \textit{in-situ} tuning between topological and trivial configurations.

To explain the basic idea, we briefly return to the case of $N=2$ JJs with phase differences $\phi_{1,2}$ and, for the moment, neglect the spin-orbit coupling\,\cite{Supplemental}. The corresponding ABS levels are shown in Fig.\,\ref{fig:4}(a) and (b). 
If the JJs are well-separated ($d\gg\xi_{S}$) and the phase differences are either identical ($\phi_{1}=\phi_{2}\equiv\phi_{*}$) or opposite ($\phi_{1}=-\phi_{2}\equiv\phi_{*}$), the ABS levels of the two JJs are degenerate. However, when the JJs are in close vicinity ($d\lesssim\xi_{S}$), the ABSs hybridize, leading to anti-crossings at the degeneracy points. At these points, the hopping amplitudes, $t(\phi_{1},\phi_{2})$, for fermionic quasiparticles are enhanced. Importantly, the hopping amplitudes for the two phase configurations are unequal, $|t(\phi_{*},\phi_{*})|\neq|t(\phi_{*},-\phi_{*})|$, as time-reversal symmetry only requires $|t(\phi_{*},\phi_{*})|=|t(-\phi_{*},-\phi_{*})|$. We will leverage this feature to implement the staggered hoppings of the SSH model. 

For simulating the SSH model, we now consider an array with $N\geq4$ JJs and repeating phase difference patterns: $\boldsymbol{\phi}_{\text{triv}}\equiv\{\phi_{*},\phi_{*},-\phi_{*},-\phi_{*},\cdots\}$ and $\boldsymbol{\phi}_{\text{topo}}\equiv\{\phi_{*},-\phi_{*},-\phi_{*},\phi_{*},\cdots\}$. These patterns are characterized by the alternating way pairs of JJs hybridize, either with equal or opposite phase differences. Using our Wannier functions approach, we derive an effective Hamiltonian for the JJ array, 
\begin{align}
\hspace{-3pt}H^{(\text{SSH})}_\mathrm{eff} = \sum_{n}\epsilon(\phi_{n}) f_{n}^\dagger f_{n}+ t(\phi_{n},\phi_{n+1})f^\dagger_{n}f_{n+1}+\text{H.c.}
\label{Eq6}
\end{align}
Similar to the case of $N=2$ JJs, when $d\gg\xi_{S}$, the ABS hybridizations are negligible, $t\approx0$, and assuming a uniform chemical potential, the on-site energies across the array are degenerate, $\{\epsilon,\epsilon,\cdots\}$, for both $\boldsymbol{\phi}_{\text{triv}}$ and $\boldsymbol{\phi}_{\text{topo}}$. In contrast, when $d\lesssim\xi_{S}$, the ABS hybridizations become significant, leading to alternating hopping strengths across the array, $\{t_{\text{in}},t_{\text{out}},t_{\text{in}}\cdots\}$. In this case, the effective Hamiltonian in Eq.\,\eqref{Eq6} simulates (up to an energy offset, $\epsilon$) the SSH model.

To verify the topological nature of our JJ array, we will now show: (1) the emergence of topological bound states and (2) the existence of a topological phase transition. For this purpose, we introduce a linear interpolation, $\boldsymbol{\phi}(\delta)$, so that $\boldsymbol{\phi}(-\phi_{*})=\boldsymbol{\phi}_{\text{triv}}$, $\boldsymbol{\phi}(0)=\bm\phi_{\text{trans}}\equiv\{0,0,\cdots\}$, and $\boldsymbol{\phi}(\phi_{*})=\boldsymbol{\phi}_{\text{topo}}$. The explicit form is: $\bm\phi(\delta)\equiv\{g_{+}(\delta),g_{-}(\delta),-g_{+}(\delta),-g_{-}(\delta),\cdots\}$ with $g_{\pm}(x)=x[h(-x)\pm h(x)]$ and $h$ the Heaviside function. With the help of Eq.\,(4), we calculate the hoppings along $\bm \phi(\delta)$. We find that for $\delta<0$, $t_{\text{out}}<t_{\text{in}}$, while for $\delta>0$, $t_{\text{out}}>t_{\text{in}}$, see Fig.\,\ref{fig:4}(d). Hence, the system undergoes a topological transition from $\bm\phi_{\text{triv}}$ to $\bm\phi_{\text{topo}}$ through $\bm\phi_{\text{trans}}$ at $\delta=0$. From calculating the spectrum in Fig.\,\ref{fig:4}(c), we indeed find that discrete subgap levels appear when $\delta>0$, corresponding to topological states at the array boundaries.

To determine the presence of the topological bound states, we propose an experiment involving the JJ array connected on both ends to normal metal leads. In this setup, local conductance measurements at each end of the array will reveal the bound state levels in the topological phase, as shown in Fig.\,\ref{fig:4}(e). Importantly, the bound state levels do not appear in the non-local conductance, depicted in Fig.\,\ref{fig:4}(f), unless there is a finite bound state hybridization. 

\textit{Conclusion.} In this work, we have introduced ABS superlattices as an all-superconducting platform for simulation lattice models of fermionic quasiparticles with gate- and phase-programmability. We introduced the formalism to derive an effective quantum simulator Hamiltonian based on Wannier functions and demonstrated the realization of topological bound states that can be probed by conductance measurements. We hope our work inspires the future exploration of ABS superlattices as an alternative quantum simulator platform.  

\textit{Acknowledgements.} We acknowledge helpful discussions with Reinhold Egger, Richard Gerhard Heß, Maximilian Hünenberger, Joel Hutchinsons, Charles. M. Marcus, Evan Thingstad, Kunal Verma, and Alexander Zazunov.

\setcounter{section}{0}
\setcounter{equation}{0}
\renewcommand*{\thesection}{S.\arabic{section}}
\renewcommand*{\thesubsection}{S.\arabic{section}.\arabic{subsection}}
\renewcommand*{\thesubsubsection}{S.\arabic{section}.\arabic{subsection}.\roman{subsubsection}}
\renewcommand{\theequation}{S.\arabic{equation}}
\setcounter{figure}{0}
\renewcommand\thefigure{S.\arabic{figure}}

\setcounter{section}{0}
\setcounter{equation}{0}
\renewcommand*{\thesection}{S.\arabic{section}}
\renewcommand*{\thesubsection}{S.\arabic{section}.\arabic{subsection}}
\renewcommand*{\thesubsubsection}{S.\arabic{section}.\arabic{subsection}.\roman{subsubsection}}
\renewcommand{\theequation}{S.\arabic{equation}}
\setcounter{figure}{0}
\renewcommand\thefigure{S.\arabic{figure}}

\newpage

\begin{widetext}

\startcontents[supplemental]

    \begin{center}
        \large{\bf Supplemental Material to `\titlecommand' \\}
    \end{center}
    \begin{center}
        Peter D. Johannsen$^{1}$\orcidlink{0000-0003-4570-8881} and Constantin Schrade$^{2}$
        \\
        {\it $^{1}$ Center for Quantum Devices, Niels Bohr Institute, University of Copenhagen, 2100 Copenhagen, Denmark}\\
        {\it $^{2}$ Hearne Institute of Theoretical Physics, Department of Physics \& Astronomy, \\ Louisiana State University, Baton Rouge LA 70803, USA}
    \end{center}
    In this Supplemental Material, we provide details on the system parameters used for the simulations in the main text. We also discuss technical aspects of the Wannier function constructions, symmetry properties of the effective Hamiltonian, and the topologial lattice model in the presence of spin-orbit coupling.
    
    \printcontents[supplemental]{}{1}{\section*{Table of Contents}}

\section{Parameters of the tight binding model}\label{SM:TBM}
    In this first section of the Supplemental Material, we provide the parameters that were used for obtaining the tight-binding results presented 
    in the main text.  The chosen parameters are thereby motivated by earlier simulations that reproduced 
    experimental measurements of Andreev bound state levels in nanowire Josephson junctions\,\cite{matute2022signatures-sm}. 
    \\
    \\
    Specifically, for our simulations, we have set: 
\begin{align}
    m^* = 0.023 m_e,\quad a_x=50\mathrm{nm},\quad a_y = 100\mathrm{nm},\quad \alpha = 22\mathrm{meV\; nm},\quad \Delta = 0.2\mathrm{meV}, \quad \lambda_0 = \frac{\hbar^2}{2m^*},
\end{align}
where $m^{*}$ is the effective electron mass, $a_x$ is the lattice constant in the $x$-direction (along the direction of the Josephson junction array), 
$a_y$ is the lattice constant in the $y$-direction (perpendicular to the direction of the Josephson junction array), $\alpha$ is a spin-orbit coupling parameter, $\Delta$ is the magnitude of the superconducting gap, and $\lambda_{0}$ is an energy scale for the hoppings.
\\
\\
With the help of these parameters, we defined,
\begin{align}
    \alpha_x = \frac{\alpha}{2a_x},\quad \alpha_y = \frac{\alpha}{a_y},\quad \lambda_{x\mathrm{SC}}=\frac{\lambda_0}{a_x^2},\quad \lambda_{x\mathrm{N}}=\frac{0.8\lambda_0}{a_x^2},\quad \lambda_{y\mathrm{SC}}=\lambda_y=\frac{t_0}{a_y^2},
\end{align}
where $\alpha_x$ is the spin-orbit coupling along the $x$-direction, $\alpha_y$ is the spin-orbit coupling along the $y$-direction, $\lambda_{x\mathrm{SC}}$ is the spin-orbit coupling along the $x$-direction in the superconducting segments, $\lambda_{x\mathrm{N}}$ is the hopping along the $x$-direction in the normal segments , and $\lambda_{y}$ is the hopping along the $y$-direction in both the normal and superconducting segments.
\\
\\
For the on-site energies, we choose,
\begin{align}
	  \mu_\mathrm{N} = 0.3\mathrm{meV},\quad \boldsymbol{\epsilon}_{\mathrm{N}} = \frac{2\lambda_0}{a_x^2}(1.2,1.08) - \mu_\mathrm{N} ,\quad \boldsymbol{\epsilon}_{\mathrm{SC}}=\left[\frac{2\lambda_0}{a_x^2}-\Delta\right](1,1)
\end{align}, 
where we assumed that the width of the system along the $y$-direction comprises $W=2$ sites. The vectors $\boldsymbol{\epsilon}_{\mathrm{N}}$ and $\boldsymbol{\epsilon}_{\mathrm{SC}}$ contain the on-site energies for these two transversal sites in the normal and superconducting regions, respectively.
\\
\\
For the remaining geometric parameters, we choose, 
\begin{align}
    N_\mathrm{O} = 20,\quad  N_\mathrm{I} = 10, \quad N_\mathrm{N}=11
\end{align}
where $N_{\mathrm{O}}$ is the number of sites in the superconducting segments at the ends of the Josephson junction array, $N_\mathrm{I}$ is the number 
of sites in the superconducting segments within the Josephson junction array, and $N_\mathrm{N}=11$ is the length of the normal segments. 
We remark that in Fig.\,3 of the main text, we set $N_\mathrm{O} = 10$, but otherwise used the geometric parameters given above. Moreover, for the Josephson junction array in Fig.\,4 of the main text, we set the number of Josephson junctions to $N=12$ and set, 
\begin{align}
    \mu_\mathrm{N} = 2.2\mathrm{meV},\quad N_\mathrm{I} = 30,\quad
    N_\mathrm{O} = 20,\quad
    \alpha=0,\quad 
    a_x=25\mathrm{nm},\quad
    a_y=50\mathrm{nm}
\end{align}
All other parameters were choosen as above. We illustrate the parameters used in the tight-binding simulations in Fig.\,\ref{fig:SNSFigure}.

\begin{figure}[H]
    \centering
    \includegraphics[width=\textwidth]{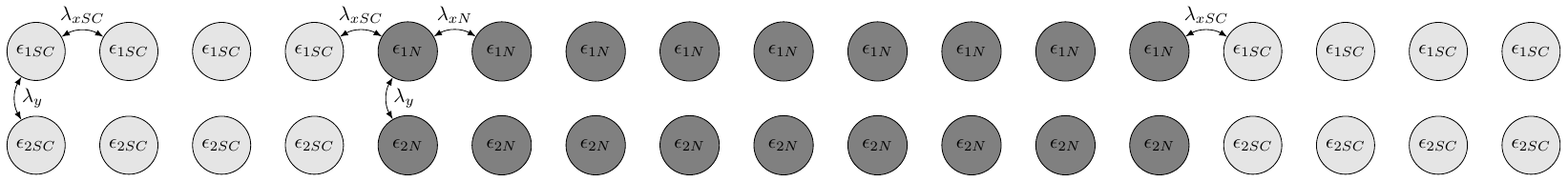}
    \caption{Illustration of the parameters used in the tight-binding simulations.}
    \label{fig:SNSFigure}
\end{figure}

\section{Construction of Wannier functions}
In this second section of the Supplemental Material, we provide details on the construction of the Wannier functions that are used for constructing 
the quantum simulator Hamiltonian of Eq.\,(3) in the main text. 
\\
\\
Conceptually, the idea of the Wannier construction is to introduce a set of localized ``trial wavefunctions" and then use a 
singular value decomposition to obtain the ``best fit" to the trial wavefunctions expressed in terms of a linear combination of eigenstates of the coupled Josephson junction system. 
\\
\\
More specifically, the singular value decomposition takes as input a matrix, 
\begin{equation}
A_{ij} = \langle e_i\vert t_j\rangle,
\end{equation}
where \{$|e_{i}\rangle$\} are the eigenstates of the coupled Josephson junction system and \{$|t_{i}\rangle$\} is a set of trial wavefunctions. The output of the singular value decomposition is a decomposition, 
\begin{equation}
A = PDQ^\dagger,
\end{equation}
where $P$ and $Q$ are unitary matrices, whereas $D$ is a diagonal matrix, consisting of the ``singular values" of $A$. 
\\
\\
Given the output data of the singular value decomposition, we construct a new matrix, 
\begin{equation}
U = PQ^\dagger = A(A^\dagger A)^{-\frac12},
\end{equation}
which is unitary by construction, and use this to change the basis from the eigenstate wavefunctions to the Wannier functions\,\cite{vanderbilt1-sm,vanderbilt2-sm}. If the eigenstate wavefunctions and the trial wavefunctions truly spanned the same subspace (which is unlikely), the singular values would all be equal to one (in which case $A$ is unitary). Removing the $D$ matrix from the decomposition ensures that the transformation from the eigenbasis to the Wannier basis is a unitary transformation, irrespective of the trial wavefunctions. However, the choice of trial wavefunctions is essential in assuring that the transformation defined by $U$ is meaningful, in the sense that it truly returns localized wavefunctions.
\\
\\
In the tight-binding simulations, we construct the trial wavefunctions for the $n^{\text{th}}$ Josephson junction site, which has a superconducting phase drop $\phi_{n}$. This is done by replacing all other Josephson junctions with superconductors, effectively creating a system with a single junction. The phase drop across this junction is assigned the value $\phi_{n}$.  We then select the trial wavefunctions as the eigenfunctions of this  single-junction system.
\\
\\
In a final step, we note that the singular value decomposition generally results in a set of Wannier functions $w_{n,\eta}(i,j)$ where $\langle w_{n,\eta}\vert H \vert w_{n,\eta'}\rangle\neq 0$ for $\eta\neq\eta'$. This means the effective spins on each Josephson junction site are coupled. To remove this coupling, we apply a unitary transformation $R_{n}$ on each site that diagonalizes the Wannier functions in the effective spin space, producing new Wannier states $w_{n,\sigma} = \sum_{\eta=u,d} R_{n,\sigma\eta} w_{n,\eta}$. In this new Wannier basis, $\langle w_{n,\sigma}\vert H \vert w_{n,\sigma'}\rangle= 0$ for $\sigma\neq\sigma'$, implying that the effective spins on each site are no longer coupled. However, this local transformation on each site also changes the hopping matrix elements, which needs to be taken into account.

\section{Anti-commutation relations for the Wannier operators}
In this third setion of the Supplemental Material, we will verify the fermionic anti-commutation relations for the constructed Wannier operators. 
\\
\\
In order to show that the Wannier operators are a set of fermionic operators, we will first show that the eigenoperators of the Hamiltonian are a fermionic set.
\subsection{Eigenoperators}
	 Let us denote the Nambu components of the eigenfunction, $\psi_n(i,j)$ as:
	\begin{align}
	  \psi_n(i,j) = \begin{pmatrix}
	\psi_{n,\uparrow}(i,j)\\
	\psi_{n,\downarrow}(i,j)\\
	\bar\psi_{n,\downarrow}(i,j)\\
	\bar\psi_{n,\uparrow}(i,j)
\end{pmatrix}
\end{align}
and let us define the corresponding eigenoperator, which can be thought of as a generalised Bogoliubov transformation of the microscopic operators:
\begin{align}
	  \hat\psi_n^\dagger = \sum_{i,j} [\psi_{n,\uparrow}(i,j) c_{i,j,\uparrow}^\dagger +\psi_{n,\downarrow}(i,j) c_{i,j,\downarrow}^\dagger +\bar\psi_{n,\downarrow}(i,j) c_{i,j,\downarrow}-\bar\psi_{n,\uparrow}(i,j) c_{i,j,\uparrow}]
\end{align}
Now, anti-commutators of the kind $\{\psi,\psi^\dagger\}$ can be related to the the single-particle wavefunctions:
\begin{align}
	  \{\psi_m,\psi_n^\dagger\} = \sum_{i,j}\sum_{k,\ell}\big[&\psi_{m,\uparrow}^\star(i,j)\psi_{n,\uparrow}(k,\ell)\{c_{i,j,\uparrow},c_{k,\ell,\uparrow}^\dagger\}+\psi_{m,\downarrow}^\star(i,j)\psi_{n,\downarrow}(k,\ell)\{c_{i,j,\downarrow},c^\dagger_{k,\ell,\downarrow}\}+\\
	  &\bar\psi_{m,\downarrow}^\star(i,j)\bar\psi_{n,\downarrow}(k,\ell)\{c_{i,j,\downarrow}^\dagger,c_{k,\ell,\downarrow}\}+\bar\psi_{m,\uparrow}^\star(i,j)\bar\psi_{n,\uparrow}(k,\ell)\{c_{i,j,\uparrow}^\dagger,c_{k,\ell,\uparrow}\}\big]\nonumber
\end{align}
Where we have left the terms that are equal to zero (these are equal to zero because of the anti-commuting properties of the microscopic operators). The (microscopic) anti-commutators above give us one iff $i=k$ and $j=\ell$. And hence we get 
\begin{align}
	  	  \{\psi_m,\psi_n^\dagger\} &= \sum_{i,j}\sum_{k,\ell}\big[\psi_{m,\uparrow}^\star(i,j)\psi_{n,\uparrow}(i,j)+\psi_{m,\downarrow}^\star(i,j)\psi_{n,\downarrow}(i,j)+\bar\psi_{m,\downarrow}^\star(i,j)\bar\psi_{n,\downarrow}(i,j)+\bar\psi_{m,\uparrow}^\star(i,j)\bar\psi_{n,\uparrow}(i,j)\big]\\
	  	  &= \langle \psi_m\vert\psi_n\rangle = \delta_{n,m}\nonumber
\end{align}
Which follows from the orthonormality of the single-particle wavefunctions. Similarly 
\begin{align}
	  \{\psi_m,\psi_n\} = -\sum_{i,j}\big[\psi^\star_{m,\uparrow}(i,j)\bar\psi^\star_{n,\uparrow}(i,j) - \psi^\star_{m,\downarrow}(i,j)\bar\psi^\star_{n,\downarrow}(i,j)-\bar\psi^\star_{m,\downarrow}(i,j)\psi^\star_{n,\downarrow}(i,j)+\bar\psi^\star_{m,\uparrow}(i,j)\psi^\star_{n,\uparrow}(i,j)\big] 
\end{align}
Which we can write as 
\begin{align}
	 \{\psi_m,\psi_n\} = \sum_{i,j}\begin{pmatrix}
	\psi^\star_{m,\uparrow}(i,j) &	\psi^\star_{m,\downarrow}(i,j) & \bar\psi^\star_{m,\downarrow}(i,j) & \bar \psi^\star_{m,\uparrow}(i,j)
\end{pmatrix}\begin{pmatrix}
	0&0&0&-1\\
	0&0&1&0\\
	0&1&0&0\\
	-1&0&0&0
\end{pmatrix}\mathcal K \begin{pmatrix}
	\psi_{n,\uparrow}(i,j)\\
	\psi_{n,\downarrow}(i,j)\\
	\bar\psi_{n,\downarrow}(i,j)\\
	\bar\psi_{n,\uparrow}(i,j)
\end{pmatrix}
\end{align}
where $\mathcal K$ is the complex-conjugation operator. The operator
\begin{align}
	  \mathcal P(i,j) = \begin{pmatrix}
	0&0&0&-1\\
	0&0&1&0\\
	0&1&0&0\\
	-1&0&0&0
\end{pmatrix}\mathcal K
\end{align}
is the particle-hole operator, which is a symmetry of the Hamiltonian and it \textit{anti}-commutes with the Hamiltonian, which implies that if $\ket{\psi}$ is an eigenstate of the Hamiltonian with energy $E$, then $\mathcal P\ket{\psi}$ is an eigenstate of the Hamiltonian with energy $-E$. Thus we have shown that $\{\psi_m,\psi_n\} = 0$ as this once again follows due to the orthonormality of the eigenbasis of the Hamiltonian. 
\subsection{Wannier operators}
From here it is a straight-forward proof to show that the Wannier operators also form a fermionic set: let 
\begin{align}
	 f_\tau = \sum_n u_n^\tau \psi_n ,\qquad f_\tau^\dagger = \sum_n (u_n^\tau)^\star \psi_n^\dagger
\end{align}
Using the fermionic properties of $\psi_n$ we get 
\begin{align}
	  \{f_\tau^\dagger,f_{\tau'}^\dagger\} &= 0\\
	  \{f_\tau,f_{\tau'}^\dagger\} &= \sum_{n} u_n^\tau  (u_n^{\tau'})^\star = \delta_{\tau\tau'}
\end{align}
where the final equality is guaranteed by the fact that the transformation the takes us from the eigenbasis to the Wannier basis is a unitary transformation.

\section{Time-reversal transformation properties}

In this fourth section of the Supplemental Material, we will summarize the transformation properties of the microscopic Hamiltonian, 
and the effective Hamiltonian  
under a time-reversal operation.

\subsection{Microscopic Hamiltonian}
We begin with the time-reversal transformation $\mathcal{T}$, which for the fermionic operators of the microscopic lattice reads,
	\begin{align}
	  \T c_{i,j,\uparrow} \T^{-1} = c_{i,j,\downarrow},\qquad \T c_{i,j,\downarrow} \T^{-1} = -c_{i,j,\uparrow},\qquad \T \kappa \T^{-1} = \kappa^\star.
\end{align}
Here, $\kappa$ is a complex number.
From these properties, it follows that the non-interacting part of the microscopic Hamiltonian satisfies,
\begin{equation}
\T H_{0}(\Phi_{n}) \T^{-1} = H_{0}(-\Phi_{n}),
\end{equation}
where (unlike in the main text) we have explicitly included the dependence of the Hamiltonian on the fluxes ($\Phi_{n}$). The equivalent relation for the Nambu space matrix representation, $\mathcal{H}_{0}$, of the Hamiltonian, $H_{0}$, is implemented as,
\begin{equation}
U^{\dag}_{T}\mathcal{H}_0^{\star}(-\Phi_{n})U_{T}=\mathcal H_0(\Phi_{n}),
\end{equation}
where $U_{T}=i\tau^0 \sigma^{y}$ and $\sigma^{y}$ denotes a spin-space Pauli matrix and $\tau^0$ denotes particle-hole-space $2\times 2$ identity. 

\subsection{Fermionic quasiparticles}
We next derive the transformation properties of the fermionic quasiparticle operators, which were defines as,  
\begin{equation}
\begin{split}
	  f_{n,\sigma}^\dagger(\Phi_{n}) &= \sum_{i,j}\; w_{n,\sigma,\uparrow}(i,j,\Phi_{n})\,c_{\uparrow,i,j}^\dagger
	  +w_{n,\sigma,\downarrow}(i,j,\Phi_{n})\,c_{\downarrow,i,j}^\dagger
	  \\
	  &\qquad+\tilde w_{n,\sigma,\downarrow}(i,j,\Phi_{n})\,c_{\downarrow,i,j}-\tilde w_{n,\sigma,\uparrow}(i,j,\Phi_{n})c_{\uparrow,i,j},
\\
\\
	  	  f_{n,\sigma}^\dagger(-\Phi_{n}) &= \sum_{i,j}\; w_{n,\sigma,\uparrow}(i,j,-\Phi_{n})\,c_{\uparrow,i,j}^\dagger
	  +w_{n,\sigma,\downarrow}(i,j,-\Phi_{n})\,c_{\downarrow,i,j}^\dagger
	  \\
	  &\qquad+\tilde w_{n,\sigma,\downarrow}(i,j,-\Phi_{n})\,c_{\downarrow,i,j}-\tilde w_{n,\sigma,\uparrow}(i,j,-\Phi_{n})c_{\uparrow,i,j}.
\end{split}
\end{equation}
\\
To these definition, we can apply the time-reversal operation,
\begin{equation}
\begin{split}
 \T f_{n,\sigma}^\dagger(\Phi_{n}) \T^{-1}   &= \sum_{i,j}\;
 -w_{n,\sigma,\downarrow}(i,j,\Phi_{n})^{\star}\,c_{\ua,i,j}^\dagger
 +w_{n,\sigma,\uparrow}(i,j,\Phi_{n})^{\star}\,c_{\da,i,j}^\dagger
	  \\
	  &\qquad
	  -\tilde w_{n,\sigma,\uparrow}(i,j,\Phi_{n})^{\star}c_{\da,i,j}
	  -\tilde w_{n,\sigma,\downarrow}(i,j,\Phi_{n})^{\star}\,c_{\ua,i,j}. 
\end{split}
\end{equation}
\\
Next, we recall that the Wannier wavefunctions were constructed to satisfy,
\begin{equation}
\begin{split}
\boldsymbol{w}_{n,d}(-\Phi_{n})^{\star}&=U_{T}\boldsymbol{w}_{n,u}(\Phi_{n}),
\\
-\boldsymbol{w}_{n,u}(-\Phi_{n})^{\star}
&=
U_{T}\boldsymbol{w}_{n,d}(\Phi_{n})
\end{split}
\end{equation}
or, equivalently,
\begin{equation}
\begin{split}
\begin{bmatrix}
 w_{n,d,\uparrow}(i,j,-\Phi_{n}) \\
  w_{n,d,\da}(i,j,-\Phi_{n}) \\
  \tilde w_{n,d,\da}(i,j,-\Phi_{n}) \\
  \tilde w_{n,d,\uparrow}(i,j,-\Phi_{n})
\end{bmatrix}
&=
\left[\begin{array}{r}
-w_{n,u,\downarrow}(i,j,\Phi_{n})^{\star} \\
w_{n,u,\ua}(i,j,\Phi_{n})^{\star} \\
-\tilde w_{n,u,\ua}(i,j,\Phi_{n})^{\star} \\
\tilde w_{n,u,\downarrow}(i,j,\Phi_{n})^{\star}
\end{array}\right].
\end{split}
\end{equation}
From these equalities, it follows that the fermionic quasiparticle obey the transformation properties,
\begin{align}
	  \T f_{n,u}^\dagger(\Phi_{n}) \T^{-1} = f_{n,d}^\dagger(-\Phi_{n}) ,\qquad \T f_{n,d}^\dagger(\Phi_{n}) \T^{-1} = -f_{n,u}^\dagger(-\Phi_{n}).
\end{align}

\subsection{On-site energies and hoppings}
We now continue by deriving properties of the on-site energies and hoppings that results from the time-reversal operation. The on-site energis and hoppings were defined as, 
\begin{equation}
\begin{split}
\epsilon_{n,\sigma}(\Phi_{n})&=
\sum_{i,j}\boldsymbol{w}_{n,\sigma}(i,j,\Phi_{n})^{\dag}\cdot
\mathcal{H}_{0}(\Phi_{n})\cdot\boldsymbol{w}_{n,\sigma}(i,j,\Phi_{n}),
\\
t_{\sigma\sigma'}(\Phi_{n})e^{i\theta_{\sigma\sigma'}(\Phi_{n})}&=
\sum_{i,j}\boldsymbol{w}_{1,\sigma}(i,j,\Phi_{n})^{\dag}\cdot
\mathcal{H}_{0}(\Phi_{n})\cdot\boldsymbol{w}_{2,\sigma'}(i,j,\Phi_{n}).
\end{split}
\end{equation}
We begin by considering the following rewriting of the on-site energies,
\begin{equation}
\begin{split}
\epsilon_{n,\sigma}(\Phi_{n})&=
\sum_{i,j}\boldsymbol{w}_{n,\sigma}(i,j,\Phi_{n})^{\dag}\cdot
\mathcal{H}_{0}(\Phi_{n})\cdot\boldsymbol{w}_{n,\sigma}(i,j,\Phi_{n})
\\
&=
\sum_{i,j}\boldsymbol{w}_{n,\sigma}(i,j,\Phi_{n})^{\dag}\cdot
U^{\dag}_{T}\mathcal{H}_{0}(-\Phi_{n})^{\star}U_{T}\cdot\boldsymbol{w}_{n,\sigma}(i,j,\Phi_{n})
\\
&=
\sum_{i,j}[U_{T}\boldsymbol{w}_{n,\sigma}(i,j,\Phi_{n})]^{\dag}\cdot
\mathcal{H}_{0}(-\Phi_{n})^{\star}\cdot U_{T}\boldsymbol{w}_{n,\sigma}(i,j,\Phi_{n})
\\
&=
\sum_{i,j}[\boldsymbol{w}_{n,\bar\sigma}(i,j,-\Phi_{n})^{\star}]^{\dag}\cdot
\mathcal{H}_{0}(-\Phi_{n})^{\star}\cdot\boldsymbol{w}_{n,\bar\sigma}(i,j,-\Phi_{n})^{\star}
\\
&=
\epsilon_{n,\bar\sigma}(-\Phi_{n})^{\star}
=
\epsilon_{n,\bar\sigma}(-\Phi_{n})
\end{split}
\end{equation}
In a similar way, we can rewrite the hoppings as,
\begin{equation}
\begin{split}
t_{\sigma\sigma'}(\Phi_{n})e^{i\theta_{\sigma\sigma'}(\Phi_{n})}&=
\sum_{i,j}\boldsymbol{w}_{1,\sigma}(i,j,\Phi_{n})^{\dag}\cdot
\mathcal{H}_{0}(\Phi_{n})\cdot\boldsymbol{w}_{2,\sigma'}(i,j,\Phi_{n})
\\
&=
\sum_{i,j}\boldsymbol{w}_{1,\sigma}(i,j,\Phi_{n})^{\dag}\cdot
U^{\dag}_{T}\mathcal{H}_{0}(-\Phi_{n})^{\star}U_{T}\cdot\boldsymbol{w}_{2,\sigma'}(i,j,\Phi_{n})
\\
&=
\sum_{i,j}[U_{T}\boldsymbol{w}_{1,\sigma}(i,j,\Phi_{n})]^{\dag}\cdot
\mathcal{H}_{0}(-\Phi_{n})^{\star}\cdot U_{T} \boldsymbol{w}_{2,\sigma'}(i,j,\Phi_{n}).
\end{split}
\end{equation}
We now distinguish between the spin-preserving hoppings and the spin-flip hoppings, which allows to make additional changes to the above expression,
\begin{equation}
\begin{split}
t_{uu}(\Phi_{n})e^{i\theta_{uu}(\Phi_{n})}
&=
\sum_{i,j}[U_{T}\boldsymbol{w}_{1,u}(i,j,\Phi_{n})]^{\dag}\cdot
\mathcal{H}_{0}(-\Phi_{n})^{\star}\cdot U_{T} \boldsymbol{w}_{2,u}(i,j,\Phi_{n})
\\
&=
\sum_{i,j}[\boldsymbol{w}_{1,d}(i,j,-\Phi_{n})^{\star}]^{\dag}\cdot
\mathcal{H}_{0}(-\Phi_{n})^{\star}\cdot  \boldsymbol{w}_{2,d}(i,j,-\Phi_{n})^{\star}
\\
&=
t_{dd}(-\Phi_{n})e^{-i\theta_{dd}(-\Phi_{n})},
\\
\text{and}
\qquad
t_{ud}(\Phi_{n})e^{i\theta_{ud}(\Phi_{n})}
&=
\sum_{i,j}[U_{T}\boldsymbol{w}_{1,u}(i,j,\Phi_{n})]^{\dag}\cdot
\mathcal{H}_{0}(-\Phi_{n})^{\star}\cdot U_{T} \boldsymbol{w}_{2,d}(i,j,\Phi_{n})
\\
&=
-
\sum_{i,j}[\boldsymbol{w}_{1,d}(i,j,-\Phi_{n})^{\star}]^{\dag}\cdot
\mathcal{H}_{0}(-\Phi_{n})^{\star}\cdot  \boldsymbol{w}_{2,u}(i,j,-\Phi_{n})^{\star}
\\
&=
-t_{du}(-\Phi_{n})e^{-i\theta_{du}(-\Phi_{n})}.
\end{split}
\end{equation}

From these rewriting of the on-site energies and the hoppings, we arrive at the following conditions, 
\begin{equation}
\begin{split}
\epsilon_{n,\sigma}(\Phi_{n}) &=   \epsilon_{n,\bar\sigma}(-\Phi_{n}),
\\
t_{uu}(\Phi_{n})e^{i\theta_{uu}(\Phi_{n})}
&=
t_{dd}(-\Phi_{n})e^{-i\theta_{dd}(-\Phi_{n})},
\\
t_{ud}(\Phi_{n})e^{i\theta_{ud}(\Phi_{n})}
&=
-
t_{du}(-\Phi_{n})e^{-i\theta_{du}(-\Phi_{n})}.
\end{split}
\end{equation}

\subsection{Effective Hamiltonian}
Lastly, we will derive the properties of the effective Hamiltonian under the time-reversal operation. The effective Hamiltonian was given by, 
\begin{equation}
    \begin{split}
    H_0^{(\mathrm{eff})}(\Phi_1,\Phi_2) &= \sum_{n\sigma} \epsilon_{n,\sigma}(\Phi_1,\Phi_2)f_{n,\sigma}^\dagger(\Phi_1,\Phi_2) f_{n,\sigma}(\Phi_1,\Phi_2) 
    \\
    &+ [t_{uu}(\Phi_1,\Phi_2)e^{i\theta_{uu}(\Phi_1,\Phi_2)}f_{1,u}^\dagger(\Phi_1,\Phi_2) f_{2,u}(\Phi_1,\Phi_2)
    \\
    & + \
    t_{dd}(\Phi_1,\Phi_2)e^{i\theta_{dd}(\Phi_1,\Phi_2)}f_{1,d}^\dagger(\Phi_1,\Phi_2) f_{2,d}(\Phi_1,\Phi_2)
    \\
    &+ \ t_{ud}(\Phi_1,\Phi_2)e^{i\theta_{ud}(\Phi_1,\Phi_2)}f_{1,u}^\dagger(\Phi_1,\Phi_2) f_{2,d}(\Phi_1,\Phi_2)
    \\
    &+ \
    t_{du}(\Phi_1,\Phi_2)e^{i\theta_{du}(\Phi_1,\Phi_2)}f_{1,d}^\dagger(\Phi_1,\Phi_2) f_{2,u}(\Phi_1,\Phi_2)+
    \mathrm{H.c.}].
    \end{split}
\end{equation}

Applying the time-reversal operation to the effective Hamiltonian yields,
\begin{equation}
    \begin{split}
    \mathcal{T}H_0^{(\mathrm{eff})}(\Phi_1,\Phi_2)\mathcal{T}^{-1} &= \sum_{n\sigma} \epsilon_{n,\sigma}(\Phi_1,\Phi_2)f_{n,\bar\sigma}^\dagger(-\Phi_1,-\Phi_2) f_{n,\bar\sigma}(-\Phi_1,-\Phi_2) 
    \\
    &+ [
    t_{dd}(\Phi_1,\Phi_2)e^{-i\theta_{dd}(\Phi_1,\Phi_2)}f_{1,u}^\dagger(-\Phi_1,-\Phi_2) f_{2,u}(-\Phi_1,-\Phi_2)
    \\
    & + \
     t_{uu}(\Phi_1,\Phi_2)e^{-i\theta_{uu}(\Phi_1,\Phi_2)}f_{1,d}^\dagger(-\Phi_1,-\Phi_2) f_{2,d}(-\Phi_1,-\Phi_2)
    \\
    &- \ 
    t_{du}(\Phi_1,\Phi_2)e^{-i\theta_{du}(\Phi_1,\Phi_2)}f_{1,u}^\dagger(-\Phi_1,-\Phi_2) f_{2,d}(-\Phi_1,-\Phi_2)
    \\
    &- \ t_{ud}(\Phi_1,\Phi_2)e^{-i\theta_{ud}(\Phi_1,\Phi_2)}f_{1,d}^\dagger(-\Phi_1,-\Phi_2) f_{2,u}(-\Phi_1,-\Phi_2)
    +
    \mathrm{H.c.}]\nonumber.
    \end{split}
\end{equation}
From this result as well as from the properties of the on-site energies and hoppings derived in the previous subsection, we conclude that the effective Hamiltonian transforms under time-reversal symmetry as, 
\begin{equation}
\mathcal{T}H_0^{(\mathrm{eff})}(\Phi_1,\Phi_2)\mathcal{T}^{-1} =     
H_0^{(\mathrm{eff})}(-\Phi_1,-\Phi_2).
\end{equation}

\section{Particle-Hole Transformation Properties}

In this fifth section of the Supplemental Material, we will summarize the transformation properties of the microscopic Hamiltonian the fermionic quasiparticle operators under particle-hole transformations. 

\subsection{Microscopic Hamiltonian}

Let us denote the particle-hole operator as $\S$, which has the following properties:
\begin{align}
    \S\; c_{\uparrow,i,j}\; \S^{-1} = -c_{\uparrow,i,j}^\dagger,\qquad \S\; c_{\da,i,j}\; \S^{-1} = c_{\da,i,j}^\dagger,\qquad \S\;\kappa\;  \S^{-1} = \kappa^\star 
\end{align}
For a single microscopic site the particle hole operator can be represented, in the Nambu basis, as 
\begin{align}
    \S = \begin{pmatrix}
        0&0&0&-1\\
        0&0&1&0\\
        0&1&0&0\\
        -1&0&0&0
    \end{pmatrix}\mathcal K
\end{align}
where $\mathcal K$ is the complex conjugation operator. The particle-hole operator can be thought of as an operator that exchanges particles with holes and vice-versa. From the properties above it follows that the non-interacting microscopic Hamiltonian satisfies
\begin{align}
    \S \; H_0\;  \S^{-1} = - H_0
\end{align}
Similarly to the time-reversal operator, we can define the Nambu-space matrix representation operator $U_{PH} = \tau_y \sigma_y$, for which it holds that 
\begin{align}
     U_{\mathrm{PH}}\; \mathcal H_0  \; U_\mathrm{PH}^\dagger = - \mathcal H_0 ^\star 
\end{align}
This relates the positive-energy wavefunctions to the negative-energy wavefunctions:
\begin{align}
    \mathcal H_0 \bm w_{\epsilon,\sigma} = \epsilon_{\epsilon,\sigma} \bm w_{\epsilon,\sigma},\qquad U_{\mathrm{PH}}\mathcal H_0 U_{\mathrm{PH}}^\dagger U_{\mathrm{PH}} \bm w_{\epsilon,\sigma} = -\mathcal H_0^\star  ( U_{\mathrm{PH}} \bm w_{\epsilon,\sigma}) =\epsilon_{\epsilon,\sigma} (U_{\mathrm{PH}} \bm w_{\epsilon,\sigma})
\end{align}

which implies that $U_{\mathrm{PH}}\bm w_\epsilon = \bm w_{-\epsilon}^\star $. 
This gives us a fundamental property of the wavefunctions:
\begin{align}
\begin{bmatrix}
w_{-\epsilon,\sigma,\ua}(i,j) \\
w_{-\epsilon,\sigma,\da}(i,j) \\
\tilde w_{-\epsilon,\sigma,\da}(i,j) \\
\tilde w_{-\epsilon,\sigma,\ua}(i,j) 
\end{bmatrix} = \left[\begin{array}{r}
    -\tilde w_{\epsilon,\sigma,\ua}(i,j)^\star \\
    \tilde w_{\epsilon,\sigma,\da}(i,j)^\star \\
    w_{\epsilon,\sigma,\da}(i,j)^\star \\
    -w_{\epsilon,\sigma,\ua}(i,j)^\star 
    \end{array}\right] \label{eq:SM:PHSym}
\end{align}
Using this property we can also relate the positive-energy creation and negative-energy annihilation operators, using $f_{\epsilon,\sigma}^\dagger = \bm w_{\epsilon,\sigma} \cdot \bm \Psi^\dagger$:
\begin{align}
    f_{\epsilon,\sigma} = (f_{\epsilon,\sigma}^\dagger)^\dagger = (\bm w_{\epsilon,\sigma}\cdot \bm \Psi ^\dagger)^\dagger = ((U_\mathrm{PH}\bm w_{-\epsilon,\sigma})^\star \cdot \bm \Psi^\dagger)^\dagger =\sum_{i,j} \left[\begin{array}{r}
    -\tilde w_{-\epsilon,\sigma,\ua}(i,j)\\
    \tilde w_{-\epsilon,\sigma,\da}(i,j)\\
    w_{-\epsilon,\sigma,\da}(i,j)\\
    -w_{-\epsilon,\sigma,\ua}(i,j)
    \end{array}\right]\cdot \left[\begin{array}{r}
        c_{i,j,\ua}\\
        c_{i,j,\da}\\
        c_{i,j,\da}^\dagger\\
        -c_{i,j,\ua}^\dagger\\
    \end{array}\right]
     = f_{-\epsilon,\sigma}^\dagger
\end{align}
where we've used that $U_\mathrm{PH}$ only has real elements. This equality, $f_{-\epsilon,\sigma}^\dagger = f_{\epsilon,\sigma}$, motivates the picture of superconductivity, \textit{the excitation picture}, which is used throughout this paper, where we neglect all negative-energy states, and only treat positive-energy excitations above the ground-state.

\section{Gauging the hopping phase}
In this sixth section of the Supplemental Material, we will explain how the 
general hopping phases in the effective Hamiltonian for our minimal two-junction system can be modified to a special form (convenient for calculations) via a gauge transformation on the Wannier functions. 
\\

We initially recall that the general form of the effective Hamiltonian,
	\begin{align}
	  H^{(\text{eff})}_{0}(\Phi_{n}) = \begin{pmatrix}
	\epsilon_{1,u}(\Phi_{n}) & 0 & t_{uu}(\Phi_{n})e^{i\theta_{uu}(\Phi_{n})} & t_{ud}(\Phi_{n})e^{i\theta_{ud}(\Phi_{n})}\\
	0 & \epsilon_{1,d}(\Phi_{n}) & t_{du}(\Phi_{n})e^{i\theta_{du}(\Phi_{n})} & t_{dd}(\Phi_{n})e^{i\theta_{dd}(\Phi_{n})}\\
	t_{uu}(\Phi_{n})e^{-i\theta_{uu}(\Phi_{n})} & t_{du}(\Phi_{n})e^{-i\theta_{du}(\Phi_{n})}&	\epsilon_{2,u}(\Phi_{n}) & 0 \\
	t_{ud}(\Phi_{n})e^{-i\theta_{ud}(\Phi_{n})} & t_{dd}(\Phi_{n})e^{-i\theta_{dd}(\Phi_{n})}&0&	\epsilon_{2,d}(\Phi_{n}) 
\end{pmatrix},
\end{align}
where we have adopted the short-hand notation, $(\Phi_{n})\equiv(\Phi_1,\Phi_2)$.
\\

This effective Hamiltonian can be rewritten in a form that manifestly obeys the time-reversal transformation properties derived in the previous section of the Supplemental Material,
	\begin{align}
	  &H^{(\text{eff})}_{0}(\Phi_{n}) =\begin{pmatrix}
	\epsilon_{1,u}(\Phi_{n}) & 0 & t_{uu}(\Phi_{n})e^{i\theta_{uu}(\Phi_{n})} & t_{ud}(\Phi_{n})e^{i\theta_{ud}(\Phi_{n})}\\
	0 & \epsilon_{1,u}(-\Phi_{n}) & -t_{ud}(-\Phi_{n})e^{-i\theta_{ud}(-\Phi_{n})} & t_{uu}(-\Phi_{n})e^{-i\theta_{uu}(-\Phi_{n})}\\
	t_{uu}(\Phi_{n})e^{-i\theta_{uu}(\Phi_{n})} & -t_{ud}(-\Phi_{n})e^{i\theta_{ud}(-\Phi_{n})}&	\epsilon_{2,u}(\Phi_{n}) & 0 \\
	t_{ud}(\Phi_{n})e^{-i\theta_{ud}(\Phi_{n})} & t_{uu}(-\Phi_{n})e^{i\theta_{dd}(-\Phi_{n})}&0&	\epsilon_{2,u}(-\Phi_{n}) 
\end{pmatrix}.
\end{align}
\\

We now introduce a unitary matrix that defines a gauge transformation on the fermionic quasiparticle operators,
\begin{align}
	 U(\Phi_{n})^{\dag} \equiv \begin{pmatrix}
	 e^{i \alpha_{1}(\Phi_{n})} & 0 & 0 & 0 \\
 0 & e^{-i \alpha_{1}(-\Phi_{n})} & 0 & 0 \\
 0 & 0 & e^{i \alpha_{2}(\Phi_{n})} & 0 \\
 0 & 0 & 0 & e^{-i \alpha_{2}(-\Phi_{n})} \\
\end{pmatrix}.
\end{align}
Here, the phases, $\alpha_{1/2}(\Phi_{n})$, are chosen so that $U(\Phi_{n})$ preserves the time-reversal transformation properties of the fermionic quasiparticle operators. More specifically, the transformed fermionic quasiparticle operators, 
\begin{equation}
\begin{split}
\tilde{f}^{\dag}_{1/2,u}(\Phi_{n})&\equiv e^{i\alpha_{1/2}(\Phi_{n})}\, f^{\dag}_{1/2,u}(\Phi_{n}),
\\
\tilde{f}^{\dag}_{1/2,d}(\Phi_{n})&\equiv e^{-i\alpha_{1/2}(-\Phi_{n})}\, f^{\dag}_{1/2,d}(\Phi_{n}),
\end{split}
\end{equation}
transform under time-reversal as,
\begin{equation}
\begin{split}
\mathcal{T}\tilde{f}^{\dag}_{1/2,u}(\Phi_{n})\mathcal{T}^{-1}
&=e^{-i\alpha_{1/2}(\Phi_{n})}\,
\mathcal{T} f^{\dag}_{1/2,u}(\Phi_{n})\mathcal{T}^{-1}
=e^{-i\alpha_{1/2}(\Phi_{n})}\,
f^{\dag}_{1/2,d}(-\Phi_{n})
=
\tilde{f}^{\dag}_{1/2,d}(-\Phi_{n}),
\\
\mathcal{T}\tilde{f}^{\dag}_{1/2,d}(\Phi_{n})\mathcal{T}^{-1}
&=e^{i\alpha_{1/2}(-\Phi_{n})}\,
\mathcal{T} f^{\dag}_{1/2,d}(\Phi_{n})\mathcal{T}^{-1}
=-e^{i\alpha_{1/2}(-\Phi_{n})}\,
f^{\dag}_{1/2,u}(-\Phi_{n})
=-
\tilde{f}^{\dag}_{1/2,u}(-\Phi_{n}).
\end{split}
\end{equation}
\\

We now adopt the choice,
\begin{equation}
\begin{split}
\alpha_{1}(\Phi_{n})&=
\frac{1}{4}
\left[
\theta_{ud}(-\Phi)
-
3
\theta_{ud}(\Phi)
-
\theta_{uu}(-\Phi)
-
\theta_{uu}(\Phi)
\right]
\\
\alpha_{2}(\Phi_{n})&=
\frac{1}{4}
\left[
\theta_{uu}(\Phi)
+
\theta_{uu}(-\Phi)
-
\theta_{ud}(\Phi)
-
\theta_{ud}(-\Phi)
\right]
\end{split}
\end{equation}
The transformed Hamiltonian then takes on the form,
	\begin{align}
	  &U(\Phi_{n})^{\dag}H^{(\text{eff})}_{0}U(\Phi_{n})
   =\begin{pmatrix}
	\epsilon_{1,u}(\Phi_{n}) & 0 & t_{uu}(\Phi_{n})e^{i\theta_{}(\Phi_{n})} & t_{ud}(\Phi_{n})\\
	0 & \epsilon_{1,u}(-\Phi_{n}) & -t_{ud}(-\Phi_{n}) & t_{uu}(-\Phi_{n})e^{-i\theta_{}(-\Phi_{n})}\\
	t_{uu}(\Phi_{n})e^{-i\theta_{}(\Phi_{n})} & -t_{ud}(-\Phi_{n})&	\epsilon_{2,u}(\Phi_{n}) & 0 \\
	t_{ud}(\Phi_{n}) & t_{uu}(-\Phi_{n})e^{i\theta_{}(-\Phi_{n})}&0&	\epsilon_{2,u}(-\Phi_{n}) 
\end{pmatrix},
\end{align}
where we have defined the new hopping phase,
\begin{equation}
\theta(\Phi_{n})\equiv
\frac{1}{2}
\left[
\theta_{ud}(-\Phi_{n})-\theta_{ud}(\Phi_{n})-\theta_{uu}(-\Phi_{n})+\theta_{uu}(\Phi_{n}),
\right]
\end{equation}
which manifestly obeys the relation, $\theta(\Phi_{n})=-\theta(-\Phi_{n})$. This phase is a gauge-invariant quantity, and therefore well-defined irrespective of the arbitrary phases chosen in numerics. 

\section{Topological lattice models with spin-orbit coupling}
In this last section of the Supplemental material, we provide more details on the topological lattice model discussed in the main text.
\\
\\
Specifically, in the main text, we focused on the case when there is no spin-splitting of the Andreev levels. Such a situation is achievable
in multiple ways: For example, if there is no spin-orbit coupling in the junction regions or the junction length is short on the scale of the spin-orbit length. Alternatively, the local on-site energies in each junction can be gate-tuned so that the Fermi velocities for the spin states are approximately equal, in which case a spin-splitting would also be absent. 
\\
\\
Nevertheless, it's still interesting to explore the case when a spin-splitting is present. In this case, the hopping parameters $t_{\sigma\sigma'}(\phi,\phi')$ are no longer spin-preserving, and the onsite energies $\epsilon_\sigma(\phi_n)$ depend on the pseudospin index. These changes yield a modified effective Hamiltonian of the form,
\begin{align}
    H^{\text{(SSH)}}_\mathrm{eff} \rightarrow \sum_{n\sigma} \epsilon_{\sigma} (\phi_n)f^\dagger_{\sigma n} f_{\sigma n} + \sum_{\sigma'}t_{\sigma\sigma'}(\phi_n,\phi_{n+1}) f_{n\sigma}^\dagger f_{n+1\sigma'} + \mathrm{H.c}.
\end{align}
Interestingly, we find that the edge modes still airse in the spin-orbit coupled case, as is illustrated in Fig.\,\ref{fig:SM:4ens}.
However, in general, we find that the edge modes are also split in energy and can be separated by bulk states. 
\begin{figure}
    \centering
    \includegraphics[width=0.6\linewidth]{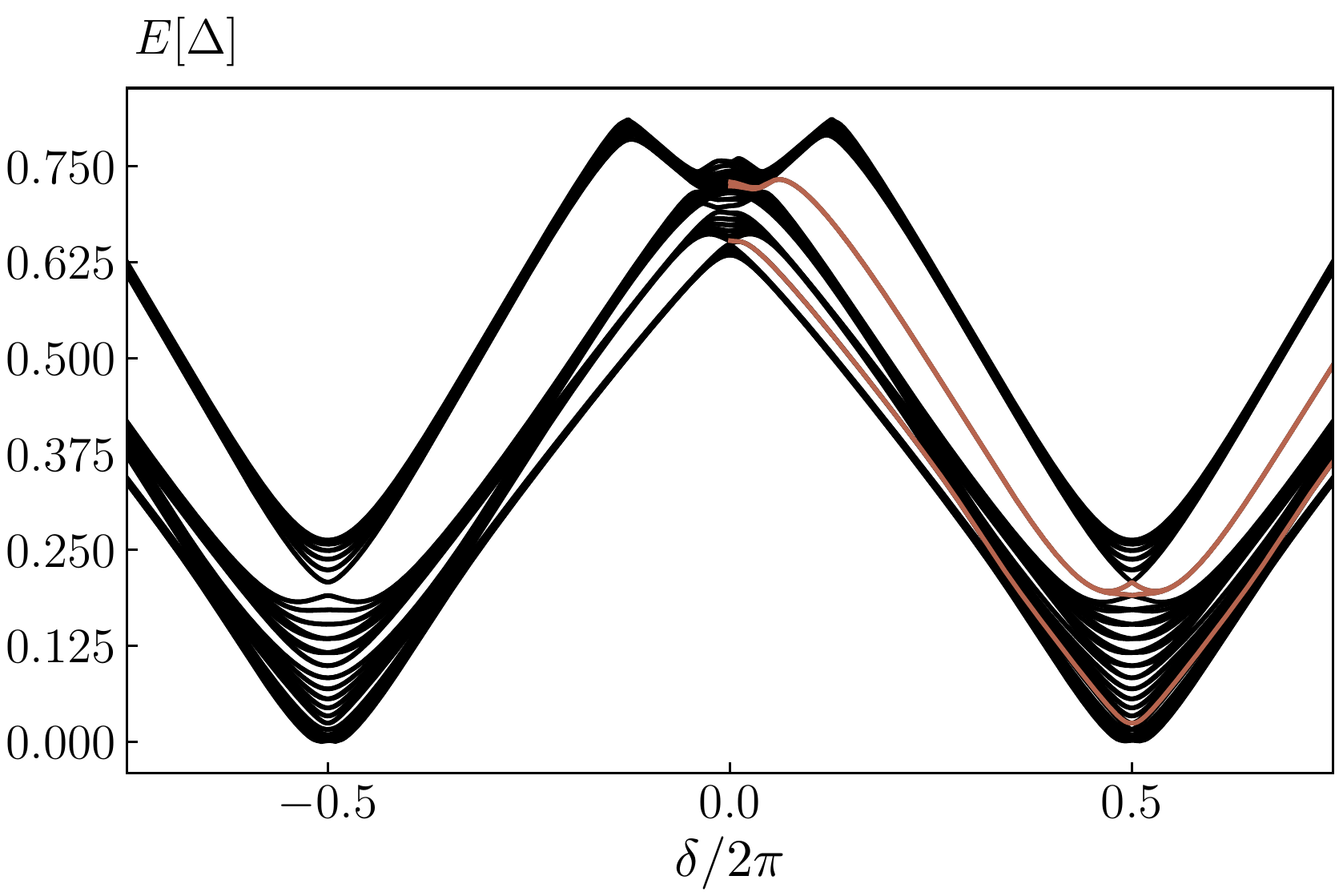}
    \caption{ABS levels versus $\delta$ for the topological lattice model in the presence of a spin-orbit-induced spin-splitting. In this case,
    the edge modes (red) still arise, but are in general spin-split. }
    \label{fig:SM:4ens}
\end{figure}

\end{widetext}

\stopcontents[supplemental]

\end{document}